\pdfoutput=1 
\documentclass[prd,twocolumn,nofootinbib,preprintnumbers,superscriptaddress,showpacs]{revtex4-1}

\usepackage{graphicx}
\usepackage{bm}
\usepackage{amsmath}
\usepackage{lineno}
\usepackage{amssymb}
\usepackage[colorlinks,pdfstartview=FitH]{hyperref}
\hypersetup{linkcolor=blue,citecolor=blue,filecolor=black,urlcolor=blue}
\newcommand\sect[1]{\emph{#1.}---}

\def \be {\begin{equation} }
\def \ee {\end{equation}}

\def \bem {\begin{multline}}
\def \eem {\end{multline}}

\def \bes {\begin{subequations} }
\def \ees {\end{subequations}}

\def \pd {\partial}

\def \phase {(2\pi)^3}

\def \e {\epsilon}

\def \o {\omega}

\def \s {\sigma}

\def \O {\Omega}

\def \G {\Gamma}

\def \<{\langle}
\def \>{\rangle}
\def \+{\dagger}
\def \({\left(}
\def \){\right)}
\def \[{\left[}
\def \]{\right]}

\def \vq {\bm{q}}
\def \vx {\bm{x}}

\def \vj {\bm{j}}

\def \vP {\bm{P}}

\def \vB {\bm{B}}
\def \vE {\bm{E}}

\def \vv {\bm{v}}

\def \anom {\chi}

\def \imp {\text{imp}}

\def \vlim {v_{\rm lim}}
\def \kz {k_{z}}
\def \vF {1}

\def \qz {q_{z}}
\def \cmw {v_{\chi}}

\usepackage[normalem]{ulem}

\begin{document}

\preprint{MIT-CTP/4737}
\author{Andrey V. Sadofyev}
\email{sadofyev@mit.edu}
\affiliation{Center for Theoretical Physics, Massachusetts Institute of Technology, Cambridge, MA 02139, USA}
\affiliation{ITEP, B. Cheremushkinskaya 25, Moscow, 117218, Russia}

\author{Yi Yin}
\email{yyin@bnl.gov}
\affiliation{Physics Department,
Brookhaven National Laboratory, Upton, NY 11973, USA}

\title{
Drag suppression in anomalous chiral media
}
\date{\today} 

\begin{abstract}
%
We study a heavy impurity moving longitudinal with the direction of an external magnetic field in an anomalous chiral medium. Such system would carry a non-dissipative current of chiral magnetic effect associated with the anomaly. 
We show, by generalizing Landau's criterion for superfluidity, that the ``anomalous component" which gives rise to the anomalous transport will {\it not} contribute to the drag experienced by an impurity. 
We argue on a very general basis that those systems with a strong magnetic field would exhibit an interesting transport phenomenon
--  the motion of the heavy impurity is frictionless, in analogy to the case of a superfluid. 
 We demonstrate and confirm our general results with two complementary examples: weakly coupled chiral fermion gases and strongly interacting chiral liquids.
\end{abstract}

\maketitle 

\section{Introduction}
Recently, novel transport phenomena of chiral (parity-violating) media tied with the quantum anomaly has attracted a lot of interest. 
A prominent example is the Chiral Magnetic Effect (CME) \cite{Vilenkin:1980fu,Kharzeev:2012ph,*Kharzeev:2013ffa},
the presence of the vector (electric) current along the direction of the magnetic field $\vB$:
\begin{equation}
\label{CME}
\vj_{V}=C_{A}\mu_{A}\vB\, .
\end{equation}
Here $C_{A}=1/2\pi^2$ is the coefficient in front of the chiral anomaly:
$\pd_{\mu}j^{\mu}_{A}
=  C_{A} \vE\cdot \vB$
and $\mu_{A}$ is the axial chemical potential.
The broad set of systems exhibiting CME and other related anomalous transport phenomena is widely discussed in the literature and includes the primordial electroweak plasma in early universe (see e.g. \cite{Giovannini:1997eg}), the QCD matter created in heavy-ion collisions 
(see e.g. \cite{Kharzeev:2007jp,*Kharzeev:2007tn,*Liao:2014ava}) and newly discovered condensed matter systems - Weyl and Dirac semimetals~(see e.g. \cite{ciudad2015weyl, semimetal}). 

One salient feature of those anomaly-induced currents is that they are dissipationless.
In this respect, a chiral medium is similar to a superfluid. 
In contrast to the Ohm current, 
both the CME current and the flow of the superfluid component do not produce entropy.
Another interesting feature of an ordinary superfluid is that the motion of a heavy impurity through a superfluid is frictionless. 
What would happen if a heavy impurity is moving in an anomalous chiral medium?

 In this paper, 
 we wish to obtain further insights into the non-dissipative nature of the anomalous transport by studying drag force acting on a heavy-impurity in a chiral media. 
Specifically,
we put those chiral systems in the presence of a background magnetic field $\vB$.
Those systems would exhibit a CME current \eqref{CME} that is originated from the ``anomalous component" (which we identify explicitly below) in the media\footnote{
As usual we assume there is no condensation which breaks the chiral symmetry.
Although the generalization of chiral effects to chiral symmetry broken phase is reported in literature, see e.g.~\cite{Lublinsky:2009wr, *Sadofyev:2010is}.	
This is beyond the scope of this paper.
}
. 
We then consider a heavy impurity moving longitudinal to the $\hat{B}$ direction. 
Will the motion of a heavy impurity experience any drag?

We shall show, 
by generalizing Landau's criterion for superfluidity~\cite{khalatnikov2000introduction},
that the ``anomalous component" which contributes to CME current does not contribute to the friction acting on the impurity.
This implies that CME current will not be destroyed by the presence of impurities.
We further argue in such a strong magnetic field that chiral media are populated with the ``anomalous component", 
these systems would exert no drag on an impurity moving parallel with $B$.
%
While frictionless motion of a heavy-impurity in a chiral medium is naturally connected to the non-dissipative nature of the anomalous transport, 
this novel phenomenon has not been reported to the extent of our knowledge. 
We emphasize that the necessary condition to realize this drag-free motion is the presence of a strong magnetic field. 
In the weak magnetic field regime as considered in Refs.~\cite{Rajagopal:2015roa, Stephanov:2015roa}, 
there will still be a non-dissipative anomalous current but drag force is non-zero in those situations.


Our general results will be illustrated and supported by two examples. 
We show that the drag force is absent for a weakly-coupled chiral fermion gas when the impurity velocity satisfies $v<1$ and for a strongly coupled chiral fluid when $v<\cmw$.
Here $\cmw$ is the speed of the chiral magnetic wave (CMW)~\cite{Kharzeev:2010gd}.
Remarkably, 
$\cmw$ will also approach $1$ in the strong magnetic field limit~\cite{Kharzeev:2010gd}, 
therefore results in those two examples coincide.
Throughout the paper, we set the Fermi velocity (or the speed of light) and the electric charge $e$ to unity. 
\vspace{0.5mm}
 
 \section{Landau's criterion for superfluidity and anomalous transport}
 Let us begin by reviewing the pertinent ingredients of Landau's criterion for superfluidity.
We consider a heavy impurity moving through a medium along $z$-direction (without loss of generality) with velocity $v$.
If this motion is accompanied by friction, 
a part of momentum $\qz$ carried by the heavy impurity will be transferred to the medium 
and the heavy impurity will lose kinetic energy $\O = v \qz +{\cal O}(\qz^2)$.
As usual we assume that $\qz$ is much smaller than the momentum of the heavy impurity $Mv$.
In other words, 
we assume that the impurity is so heavy that it takes many collisions to change its momentum substantially. 
This is the condition that motion of the heavy impurity can be described as a random walk and has been widely used in previous studies (cf.~Ref.~\cite{Moore:2004tg}).
Meanwhile, by gaining momentum $\qz$, the energy of the medium will increase by the amount $\Delta E(\qz)$ which equals to $\O$ by the conservation of energy. 
This transition will be possible only if 
\begin{equation}
\label{LC}
\Delta E (\qz)/\qz=\O/\qz  = v\, . 
\end{equation}
Thus, if $\Delta E(\qz)/\qz$ has a finite minimum, $v_{\rm lim}$, 
the kinematic constraint \eqref{LC} cannot be satisfied for $v<\vlim$ and consequently the friction would be forbidden. 
$\vlim$ is sometimes referred as ``critical velocity'' or ``limiting velocity''. 
For a Bose-liquid at zero temperature,
the system is in the ground state.
Therefore $\Delta E(\qz)$ and $v_{\rm lim}$ are directly related to the excitation spectrum of the system. 
This is a familiar picture about the ordinary superfluid.
It is worth pointing out here that the applicability of Landau's criterion for superfluidity, 
which relates the vanishing of the drag force to the specific features of the spectrum of the media, is not limited to the superfluid. 
Furthermore, the drag suppression also does not necessarily imply that the medium is a superfluid, 
see Refs.~\cite{bosedrag,2014arXiv1407.6184B,*Bobrov2012,*2013PTEP.2013d3I01B} and references therein for examples. 

We now turn to an anomalous chiral medium, the central topic of this paper. 
Let us first consider the simplest example of such a system: a gas of right-handed chiral fermions.
In the presence of a static and homogeneous magnetic field $\vB$ along,
say $z$-direction,
the energy spectrum of those chiral fermions is given by Landau levels (LL) \cite{landau1977quantum}.
The higher Landau levels (HLL) are even with respect to the momentum along $z$-direction $\kz$, 
i.e., $E_{n}(\kz)=\sqrt{2n B + \kz^2}$ for $ n=1,2,3\ldots~$.
In contrast, 
the lowest Landau Level (LLL) is odd in $\kz$ and it has a level crossing point at $\kz=0$:
\begin{eqnarray}
\label{E_LLL}
E_{\rm LLL}(\kz)=\kz\, .
\end{eqnarray}
%
It is well known that 
HLL will not contribute to the CME current $\vj_{z}$ as the contribution from a fermion with $\kz$ cancels that with $-\kz$~\cite{Metlitski:2005pr}.  
On the other hand, 
the contribution from LLL is non-zero and precisely gives CME current in \eqref{CME}.
%
We therefore identify LLL as the ``anomalous component" and HLL as the ``normal component" of the system. 

Let us consider the motion of a heavy impurity along $\hat{B}$. 
We see immediately, by extending Landau's criterion that LLL would not contribute to the friction. 
Indeed, 
if a chiral fermion at LLL gains momentum $\qz$ from the impurity, 
the energy of that chiral fermion increases, according to the dispersion relation of LLL: $E_{\rm LLL}(\kz)=\kz$, by the amount $\Delta E= \qz$, 
i.e. $\Delta E/\qz = 1$. 
Obviously, the kinematic constraint \eqref{LC} will not be satisfied for a heavy-impurity moving at $v<\vF$.
One can show similarly that the transition from LLL to HLL is also energetically impossible for $v<\vF$. 

The fact that LLL is chiral 
plays a crucial role in this analysis.
For a dispersion relation which is even with respect to $\kz$, 
the kinematic constraint can always be satisfied via backward scattering.
This difference is intuitively clear:
when colliding with heavy particles,
light particles will be bounced back.
However, 
such backward scattering is forbidden for chiral fermions at LLL as 
their chirality is slaved to the direction of their momentum along $z$-direction.
Therefore, those chiral fermions will pass through the heavy particle without transferring momentum. 

One could extend the analysis above to other chiral media in the presence of 
magnetic field
$\vB$. 
In those systems, 
there are zero modes, similar to LLL, in the background of a magnetic flux tube with a level-crossing point $\e_{\rm zero}(\kz=0)=0$ (see e.g. Ref.~\cite{Metlitski:2005pr} and references therein).
The existence of such zero modes is essential for the realization of the chiral anomaly and CME~\cite{Nielsen}.
Since those zero modes are  ${\cal P}$-odd, 
expanding their dispersion relation around the level crossing point, 
one will always find $\e_{\rm zero}(\kz)\propto \kz$ which is chiral. 
Therefore, the existence of the zero modes with chiral dispersion relation should be universal, independent of the microscopic details of the system~\footnote{
For systems with size smaller than $1/\sqrt{B}$, zero modes
would be absent and we do not consider this possibility.
}.
Consequently, 
our previous analysis on LLL can be readily applied to zero modes of other chiral media with the anomaly. 

The concept of zero modes is of topological nature.
Hence it is well-expected that  they would not contribute to the dissipation. 
In fact, the dissipationless nature of the anomalous transport has been used as a guiding principle in Refs.~\cite{Son:2009tf,Kharzeev:2011ds} to derive the anomalous hydrodynamics. 
Despite this,
a simple {\it microscopic} physical interpretation presented in this paper, 
which connects the chiral dispersion of zero modes and Landau's criterion for superfluidity, has not been revealed till now.
\vspace{0.5mm}

\section{Drag suppression in chiral media}
We now consider a chiral medium at finite density and temperature $T$ in a strong external magnetic field $eB\gg \mu^2, T^2$ that the medium is dominant by the zero modes.
For such systems with one chirality (i.e. with right or left-handed fermions only),
it is clear that a heavy impurity moving longitudinally to $\hat{B}$ will not experience any drag force for $v$ smaller than a characteristic value, 
determined by the chiral dispersion of the zero modes.
In analogy to the ordinary superfluid that the scarcity of low lying excited states implies the ``superfluidity'', 
the lack of ``non-zero modes" for chiral media in the strong magnetic field limit is the physical origin of the drag suppression.

The situation with more than one chirality, say with both right-handed $R$ and left-handed $L$ fermions, is however more subtle.  
When the medium-impurity coupling is weak, 
the dominant process for energy transfer involves only one chirality,
e.g. the two-to-two scattering such as $HL\to HL$ and  $HR\to HR$  where $H$ denotes the heavy impurity. 
Thus, our previous conclusion on the absence of drag force still holds. 
Of course, 
there are multi-scattering processes
which satisfy the kinematic constraint \eqref{LC}.
Nevertheless, 
this type of process will be suppressed as far as the coupling between the impurity  and the medium is weak. 
Indeed, 
there are examples that Landau's criterion is not applicable if the medium-impurity interaction is not weak (cf.~Refs.~\cite{bosedrag}).  
%

In general, 
the chirality flipping rate $\G_{\chi}$ is non-zero yet $1/\G_{\chi}$ is typically much longer than the characteristic time scales of other relaxation processes. 
In a Weyl semimetal, 
the inverse of $\G_{\chi}$ is given by the inter-valley scattering mean free time while 
in quark-gluon plasma, 
$\G_{\chi}$ is linked to the sphaleron transition rate.
With finite $\G_{\chi}$, 
there could be scattering processes $HL \to HR $ in which $L$ gains momentum and becomes $R$
that could satisfy the kinematic constraint \eqref{LC}. 
Nevertheless, the probability of those processes should be suppressed when $\G_{\chi}$ is small.
We will discuss the effects of finite $\G_{\chi}$ later.
\section{Examples} 
We now support our general results by the explicit computation of the drag force.
Below, 
we will assume the coupling between the medium and impurity is weak, 
but the interaction among chiral fermions can be strong. 

Let us start with a generic interacting Hamiltonian between the fermion density $n_{V}(\vx)$ and the
impurity density $n_{\imp}(\vx)$:
\begin{equation}
  \label{eq:HI}
  H_{I} =  \int d^3\vx\int d^3\vx'\,  n_{V}(\vx, t) U(\vx-\vx') n_{\imp}(\vx', t)\, , 
\end{equation}
where $U(\vx-\vx')$ is the interacting potential.
The momentum transfer rate between the impurity and the medium, 
at the leading order in the medium-impurity coupling (but to {\it all orders} in medium-medium couplings), 
can be derived using Fermi's golden rule~\cite{Bellac:2011kqa,nozieres1999theory}:
\begin{eqnarray}
  \label{eq:Gamma_rate}
 \phase\frac{d {\cal R}}{d^3\vq}  &=&
|U_{\vq}|^2 \frac{ \rho(\o, \vq)}{1- e^{-\o/T}} \, . 
\end{eqnarray}
Here $U_{\vq}$ is the Fourier transform of $U(\vx)$ and $\rho(\o, \vq)$ is the spectral density. 
As usual,
$\rho(\o, \vq)=-2{\rm Im}G^{R}(\o, \vq)$ with $G^{R}(\o,\vq)\sim \<n_{V}n_{V}\>$ the retarded Green's function. 
The drag force is related to the momentum transfer rate: ${\bm F}=d\vP/dt =-\int_{\vq}{\cal R}(\O_{\vq},\vq)\vq$
where $\int_{\vq} \equiv d^3\vq/\phase$. 
By noting $\rho(\o, \vq)=-\rho(-\o, -\vq)$,
we then have:
\begin{eqnarray}
\label{F}
{\bm F}
&=&\frac{d\vP}{d t}
=- \int_{\vq} \vq |U_{\vq}|^2 \rho(\O=\vv\cdot\vq,\vq)
\, . 
\end{eqnarray}
The general expression \eqref{F} has been widely used in studying the drag force in the condensed-matter literature (c.f~\cite{nozieres1999theory}). 
\eqref{F} can also be matched to the computation of the drag force in quark-gluon plasma via perturbation theory~\cite{Moore:2004tg,*CaronHuot:2007gq}.
%
 %

As the first example, 
we consider weakly coupled chiral fermions in the presence of $\vB$. 
$\rho(\o,\vq)$ can be computed by considering a polarization loop with fermion propagator in the presence of magnetic field.
In strong $B$ limit, only LLL contributions are needed~\cite{Gorbar:2011ya,*Gorbar:2013dha,*Fukushima:2011nu}:
\begin{equation}
\label{rho_LLL}
\rho_{LLL}(\o,\vq) = \(\frac{eB}{2\pi}\) e^{-\vq^2_{\perp}/2 B^2} \rho_{2D}(\o,\qz)\, ,
\end{equation}
where $\rho_{2D}(\o, \qz)$ denotes the spectral density for a free chiral-fermion in $1+1$ dimension (2D) and ${\bm q}_{\perp}$ denotes the momentum transverse to $\hat{B}$.
\eqref{rho_LLL} is expected from the dimensional reduction. 
 Since 2D anomaly receives no thermal corrections, $\rho_{2D}(\o, \qz)$ is independent of temperature and density~\cite{Baier:1991gg}: 
\begin{eqnarray}
\label{rho_2D}
\rho_{2D}(\o, \vq)
&= &
-\frac{2}{\pi}{\rm Im}\[\frac{\o^2-\qz^2}{\(\o+i\epsilon\)^2-\qz^2}\]
\nonumber \\
&=&
\o
\[\delta(\o- \qz)+\delta(\o+\qz)\]\, . 
\end{eqnarray}
As expected, 
the delta function in \eqref{rho_2D} is related to the chiral dispersion of LLL with $\delta(\o\pm\qz)$ corresponding to right-handed and left-handed chiral fermions.
It is clear by substituting \eqref{rho_2D}, \eqref{rho_LLL} into \eqref{F} that for $v_{z}<\vF$ the drag force
$F_{z}$ vanishes due to the delta function in \eqref{rho_2D}. 
This manifests the kinematic constraint as we discussed earlier. 
Note that if the heavy impurity is moving along the direction transverse to $\hat{B}$, 
 drag force is in general non-zero~\cite{Fukushima:2015wck}. 
 \vspace{0.5mm}

As another example, 
we now consider a strongly interacting chiral fluid in the presence of a background magnetic field $\vB$ in which the mean free path is small.
We could then assume that characteristic momentum transfer $\vq$ between the impurity and the medium is in the hydrodynamic regime.
Consequently,
the form of $G^{R}(\o,\vq)$ as well as $\rho(\o, \vq)$ will be completely fixed by hydrodynamics~\cite{Kadanoff1963419}.

We start with the constitute relation for the vector and axial currents $\vj_{V}, \vj_{A}$.
The CME current \eqref{CME} and its cousin in the axial current - the charge separation effect~\cite{Son:2004tq} ($\vj_{A}\sim C_{A}\mu_{V}\vB$) modifies the constitute relation~(see e.g. \cite{Newman:2005hd,Son:2009tf}):
\begin{eqnarray}
\label{eq:constituteV}
J^{l}_{V}&=& C_{A}\mu_{A} B^{l} + \s^{lm} E_{m}\, -D^{l m}\nabla_{m}n_{V}\, ,
\\
\label{eq:constituteA}
J^{l}_{A}&=& C_{A}\mu_{V}B^{l} -D ^{lm}\nabla_{m}n_{A}\, , \,\,\,\,  l,m=1,2,3\, . 
\end{eqnarray}
where $D^{ij}$ is the diffusion coefficient tensor which is related to the susceptibility $\chi$ and the conductivity tensor $\s^{ij}$ by Einstein's relation $\sigma^{ij}=\chi D^{ij}$.
 $D^{ij}$ may be decomposed \cite{lifshitz1981physical} as
$
D^{ij} = D_{T} ( \delta^{ij}-\hat{B} ^{i}\hat{B}^{j}) +
D_{L} \hat{B} ^{i}\hat{B}^{j} 
$. 
We similarly introduce the transverse and longitudinal conductivities $\s_{T,L}=\chi D_{T,L}$.
We also note that the constitute relation depends on the choice of fluid frame.
In \eqref{eq:constituteV}, \eqref{eq:constituteA}, we will use the ``no-drag'' frame, 
the frame in which the drag force will be absent if the impurity is at rest, that has been recently discussed in Refs.~\cite{Rajagopal:2015roa, Stephanov:2015roa}. 

We now restrict our discussion to a neutral chiral fluid, i.e. background $\mu_{V, A}=0$.
This also allows us to neglect possible axial current induced by electric field~\cite{Huang:2013iia} and additional non-linear term in $B$ as considered in Ref.~\cite{Gorbar:2016qfh}.
and perturb the system by imposing a space-time dependent gauge potential $\delta A_{0} \propto e^{- i \o t + i \vq\cdot\vx}$. 
That perturbation will lead to currents and chemical potentials $ \delta \mu_{V,A}$ fluctuations. 

Substituting \eqref{eq:constituteV}, \eqref{eq:constituteA} into 
$\pd_{\mu}j^{\mu}_{V}=0$ and $\pd_{\mu}j^{\mu}_{A}=C_{A}\vE\cdot \vB -\G_{\chi}n_{A}$ where we have incorporated the finite chirality flipping rate $\G_{\chi}$,
we have:
\begin{equation}
\label{eq:chargeVA}
\(
\begin{array}{cc}
\o+i \G_{\vq} ,\,& -v_{\chi}\qz\\
-v_{\chi}\qz,\,& \o+i \G_{\vq}+i\G_{\chi}
\end{array}
\)
\(
\begin{array}{c}
  \delta n_V\\
  \delta n_A
\end{array}
\) 
=
\chi
\(
\begin{array}{c}
i\G_{\vq}\\
 -v_{\chi}\qz
\end{array}
\)
 \delta A_{0}
\, .
 \end{equation}
 Here,
 we use the linearized equation of state $\delta n_{V,A}=\chi \delta\mu_{V, A}+O(\delta\mu^2)$ and introduce
\begin{equation}
\label{G_CMW}
\Gamma_{\vq}\equiv q_{i}D^{ij}q_{j}=\( D_{L} q^2_{z}+ D_{T}q^2_{\perp}\)\, ,
\,\,\,\,\,
\cmw \equiv \( C_{A} B\)/\chi\, . 
\end{equation}
One immediately finds that $\cmw$ is the speed of CMW~\cite{Kharzeev:2010gd}.

Before proceeding further, 
let us discuss  $\cmw$ and $D_{T}, D_{L}$ in large $B$ limit.
As shown in Ref.~\cite{Kharzeev:2010gd},
 $\cmw$ approaches $\vF$ in large $B$ limit. 
This also implies that $\chi\to C_{A}B$ in this limit.
Under the Drude approximation, 
$\s_{T}\sim 1/B^2$ as the motion of charge carriers in transverse plane is suppressed by large $B$. 
Meanwhile, $\s_{L}$ shows no dependence on $B$ as charge carriers moving along $\hat{B}$ feels no Lorentz force. 
Such $B$-dependence has also been observed in holographic models~\cite{Ammon:2009jt}.
By noting $D_{T,L}=\chi^{-1}\s_{T,L}$
we will take
\begin{equation}
\label{large_B}
 \cmw \to \vF\, , 
\, \, \,
\chi \to C_{A}B\, , 
 \, \, \,
D_{L} \sim B^{-1}\, ,
\, \, \,
D_{T}\sim B^{-3}\, ,
\end{equation}
for the discussion below. 
 In this limit, we could neglect $D_{T}$ dependence of $\G_{\vq}$ and make approximation $\G_{\vq}\approx \G_{\qz}\equiv D_{L}\qz^2$.  

We now read the relation between $\delta n_{A}$ and $\delta A_{0}$ from \eqref{eq:chargeVA} and match the results to the definition of the retarded Green function:
$\delta n_{V}= G_{R}(\o, \vq)\,\delta A_{0}$. 
This gives (see also Ref.~\cite{Yin:2013kya}):
\begin{eqnarray}
\label{G_hydro}
G_{R}(\o, \qz)
&=& \chi \[1- \frac{\o \(\o+i\,\Gamma_{\qz}+i\Gamma_{\chi}\)}{\Delta(\o, \qz)} \]\, ,
\\
\Delta(\o, \qz) 
&=& \(\o+i\Gamma_{\qz}\)\(\o+i\,\Gamma_{\qz}+i\G_{\chi} \)-\cmw^2\qz^2\, ,
\end{eqnarray}
where due to \eqref{large_B}, $\vq_{\perp}$-dependence is suppressed.
$\Delta(\o,\qz)=0$ determines the dispersion relation of collective excitations of the system.
We first consider the situation that $\G_{\qz}\gg \G_{\chi}$.
We then have
$\o(\qz)=\pm v_{\chi} q_{z}- i \G_{\qz}\,$. 
This is nothing but CMW~\cite{Newman:2005hd, Kharzeev:2010gd}
and $\rho=-2{\textrm Im} G_{R}$ gives:
\begin{eqnarray}
\label{rho_hydro}
\rho_{\rm hyd}(\o, \qz)
&=&  
\[\frac{\o\chi \G_{\qz}}{\(\o-v_{\chi}q_{z}\)^2+\G^2_{\qz}}+\(v_{\chi}\to -v_{\chi}\)\]
\, .
\end{eqnarray}

As a result of \eqref{large_B}, \eqref{rho_hydro} becomes:
\begin{eqnarray}
\label{rho_hydroB}
\rho_{_{\rm hyd}}(\o, \vq)
%
&\approx&
\o \chi\pi 
\[\delta(\o-v_{\chi}q_{z})+\delta(\o+v_{\chi}q_{z})\]
\, .
\end{eqnarray}
Again, 
the drag force vanishes for $v<\cmw$.
In contrast to the previous example, 
for the strongly interacting chiral fluid,
the energy transfer is not related to elementary excitations but, instead, to the collective excitation - CMW.
However, 
the absence of drag is still robust.
\vspace{0.5mm}

\begin{figure}
\centering
\includegraphics[width=0.4\textwidth] {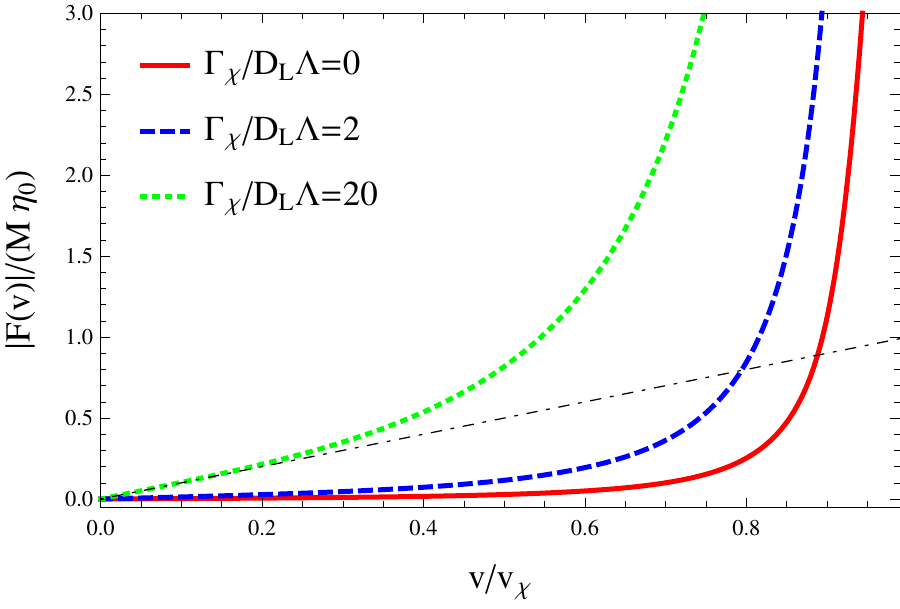}
\caption
{
\label{fig:drag_force}
Drag force $F(v)$ of a heavy impurity moving along $\hat{B}$-direction with velocity $v$ in an anomalous chiral fluid with a strong magnetic field.
We employ a toy potential (see text for details).
Dash-dotted black curve plots $M \eta_{0} v$. 
Here, 
we introduced $\eta_{0}$, which is $\eta$ 
with $\G_{\chi}/D_{L}\Lambda = 20$ to normalize $F(v)$.
}
\end{figure}
\subsection{Effects of chirality flipping rate}
%
We now return to the effects of a finite chirality flipping rate $\G_{\chi}$ using \eqref{G_hydro}. 
It is instructive to first take the opposite limit that $ \G_{\chi}\ll \G_{\qz}$.
The dispersion relation from $\Delta(\o, \qz)=0$ then leads to the ``anomalous" diffusive mode~\cite{Landsteiner:2014vua,*Stephanov:2014dma}:
\begin{equation}
\label{D_anom}
\o(\qz) = -i D_{\anom}\qz^2\, ,
\qquad
D_{\anom}=  \cmw^2/\G_{\chi}\, ,
\end{equation}
where $D_{\chi}$ is directly related to the anomaly-induced negative magnetoresistance $\s_{\chi}$~\cite{PhysRevB.88.104412} via the Einstein relation $\s_{\chi}=\chi D_{\chi}$. 
The behavior of $\rho$ is then dominated by this diffusive mode. 
The drag force now becomes non-zero even for a very small $v$. 
However, the drag force coefficient $\eta$, which enters in $F(v)=-\eta M v$ for small $v$ is tied to the chirality flipping rate $\G_{\chi}$. 
Indeed, 
for $v\ll \cmw$, 
$\rho(\o,\vq)$ obtained from \eqref{G_hydro} gives
$\rho(\o=v q_{z}, \qz)= ( 2 v \,\chi\, \G_{\chi})/(\cmw^2\,q_{z})$.
Substituting it into \eqref{F}, 
we have $\eta \propto \Gamma_{\chi}$.
%
%
%
%
%

To illustrate the effects due to intermediate $\G_{\chi}$, i.e. $\G_{\chi}\sim \G_{q_{z}}$ and summarize the above two sections, 
we compute the drag force $F(v)$ using a toy potential $U(\vq)$.
We consider a potential of the form : $U(\vq)= U_{\perp}(\vq_{\perp}) e^{-\qz^2/\Lambda^2}$ where $\Lambda$ sets the cut-off scale for hydrodynamic regime. 
$F(v)$ is computed using \eqref{F}, \eqref{G_hydro} with different $\G_{\chi}\ll \Lambda$. 
In numerics, we set $D_{L}\Lambda^2 = 0.001$ to mimic large $B$ limit \eqref{large_B}. 
As Fig~\ref{fig:drag_force} demonstrates, 
the suppression of the drag force is transparent for small $\G_{\chi}$.
For sufficiently large $\G_{\chi}$,
$F(v)$ grows linearly for small $v$ with drag force coefficient $\eta$ proportional to $\G_{\chi}$. 

\section{Conclusions and Outlook}%
In this paper, we observe that Landau's criterion for superfluidity can reveal the microscopic origin of the non-dissipative nature of the anomalous transport: energy and momentum exchange between an heavy impurity and zero modes is kinematically forbidden. 
We show that an anomalous chiral medium in the strong magnetic field $\vB$ limit exhibits a novel transport phenomena -- a heavy impurity moving longitudinal to $\hat{B}$ becomes frictionless.
It is worth noting that this generic result is insensitive to the specific realization of the chiral medium and particular types of impurities. 
For example, 
in a recent publication.~\cite{Fukushima:2015wck}, 
the heavy quark drag force coefficient of quark gluon plasma (QGP) with strong magnetic field has been computed at the leading order in $\alpha_{s}$, where $\alpha_{s}$  is the strong coupling constant. 
The suppression of drag force longitudinal to the magnetic field direction has also been observed. 

We now discuss potential applications of our results. 
A very strong magnetic field is present at early times of the QGP created in heavy-ion collisions. 
The suppression of longitudinal drag will lead to a strong anisotropy in the drag force coefficient. 
As discussed in detail in Ref.~\cite{Fukushima:2015wck}, the motion of heavy quarks in QGP and thus D-meson spectrum as measured in experiments will be influenced by such anisotropy. 
%
In addition, 
it would also be interesting to explore the consequence of the drag suppression in chiral medium in condensed matter systems such as  
Weyl and Dirac semimetals. 

In the work, 
we consider the situation that the coupling of the impurity to the medium is weak. 
It would be interesting to extend the current analysis to the case when the impurity is strongly coupled to the medium. 
We defer this for the future study. 
%
%
\vspace{1mm}

\sect{Acknowledgments}
%
%
%
The authors would like to thank K.~Rajagopal, M.~Stephanov, Naoki Yamamoto for very helpful comments on the draft and K.~Fukushima, J. Goldstone, K.~Hattori, R. Jackiw, D.~Kharzeev, L.~McLerran, F. Wilczek, Ho-Ung.~Yee, V.I. Zakharov for useful conversations.
We also thank the referees for providing constructive comments and help in improving the presentations of this paper.
The work of YY was supported by DOE Contract No.~DE-SC0012704.  The part of the work of AS (in sections III and IV) was supported by Russian Science Foundation Grant No 16-12-10059, the remaining part of the work of AS was supported by DOE Contract No.~DE-SC0011090.

\bibliography{limit_v}

\end{document}